\documentclass[twocolumn,showpacs,showkeys,amsmath,amssymb]{revtex4}
\usepackage{graphicx}
\usepackage{dcolumn}
\usepackage{bm}

\begin{document}

\title{IMF isotopic properties in semi-peripheral collisions at 
Fermi energies}

\author{R.Lionti, V.Baran\footnote{\uppercase{O}n leave from
\uppercase{HH}-\uppercase{NIPNE} and \uppercase{U}niversity of 
\uppercase{B}ucharest, \uppercase{R}omania.}, M.Colonna, M. Di Toro}

\affiliation{Laboratori Nazionali del Sud INFN, Phys.Astron. Dept. 
 Catania University\\
Via S. Sofia 62, \\ 
I-95123 Catania, Italy\\ 
E-mail: ditoro@lns.infn.it}


\begin{abstract}
{We study the neutron and proton dynamical behavior along the fragmentation
path in semi-peripheral collisions: $^{58}$Fe+$^{58}$Fe (charge asymmetric, 
N/Z = 1.23) and $^{58}$Ni+$^{58}$Ni (charge symmetric, N/Z = 1.07), at 
47 AMeV.  
We observe
 that isospin dynamics processes take place also in the charge-symmetric 
system $^{58}$Ni+$^{58}$Ni, that may produce more asymmetric fragments. 
A neutron enrichment of the neck fragments is observed, 
resulting from the interplay between  
pre-equilibrium emission and the phenomenon of \textit{isospin-migration}. 
Both effects depend on the $EoS$ (Equation of State) symmetry term. 
This point is illustrated by comparing the results obtained with 
two different choices of the symmetry energy density dependence.
 New correlation observables are suggested, to study the reaction mechanism
 and the isospin dynamics.} 
\end{abstract}


\maketitle

Collisions between heavy ions with large isospin asymmetries, made 
possible by the 
recent radioactive beam developments, represent a very efficient  way 
to probe the 
structure of nuclear-$EoS$ (Equation of State) symmetry term. 
In particular the symmetry energy behavior is
influencing very dissipative reaction mechanisms, such as fragmentation 
processes, leading to important effects on fragment composition.

In central heavy ion collisions at intermediate energies the
spinodal decomposition has been proposed as a possible mechanism for fragment
formation, see Ref.\cite{PR} for a recent review.  
According to this description, fragments should reflect the 
properties of the low density phase, where they are formed. 
In charge asymmetric systems, the isospin distillation, i.e. the formation of
more symmetric fragments surrounded by a neutron richer dilute phase, 
takes place. 
Here we will focus on fragmentation in semi-peripheral collisions, where
intermediate mass fragments ($IMF$) 
are mostly produced in the overlap zone (the $neck~region$) between 
projectile-like and target-like fragments ($PLF/TLF$, the $spectator~region$),
\cite{BertschPRC17,ColonnaNPA541,SobotkaPRC50,SobotkaPRC55,
ColonnaNPA589,baranNPA730}
\cite{CasiniPRL71,Mangiarotti,MontoyaPRL73,TokePRL75,LukasikPRC55,LefortNPA662,BocageNPA676,milazzo,GingrasPRC65,DavinPRC65,ColinPRC67,PaganoNPA734}. 
We will discuss the influence of the symmetry energy on the features of 
this fragmentation mechanism. The presence of a density
gradient between the neck (low density) and the spectator (high
density) regions affects the $N/Z$ of fragments 
in a different way with respect to the spinodal decomposition mechanism
 \cite{ditoro,baran,baranPR}.
Moreover since the isoscalar density gradients are ruling the isospin 
transfer through the density dependence of the symmetry energy, we see
that measurements of isospin observables in semiperipheral collisions
will directly probe the slope of the symmetry term around saturation,
 of large importance for the structure of neutron-rich nuclei
 \cite{FurnstNPA732}.

We consider the reactions $^{58}$Fe + $^{58}$Fe 
(charge asymmetric $N/Z = 1.23$)
and $^{58}$Ni + $^{58}$Ni (charge symmetric $N/Z = 1.07$).
In fact in these collisions, due to the uniform $N/Z$ distributions, we do not 
have isospin gradients initially. We show that the study of the 
full reaction dynamics and of
the possible occurrence of density gradients is essential in order 
to understand
the isospin dynamics. We will finally suggest the measurement of
some correlations between $IMF$ properties, like mass, isospin content
and ``alignement'', particularly sensitive to the reaction mechanism and
the isospin transport.

\begin{figure}[t]
\begin{center}
\includegraphics[scale=0.50]{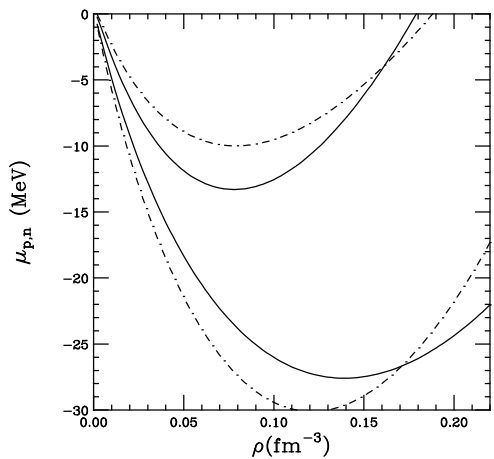}
\caption{Density dependence of the chemical potential for neutrons 
(upper curves) and protons (lower curves) for an asy-stiff (solid lines) 
and asy-soft (dashed lines) $EoS$ with asymmetry parameter $I=0.2$.}
\label{chem}
\end{center}
\end{figure}

To get an insight into the behavior of neutrons and protons 
in asymmetric matter
one can consider the density dependence of neutron and proton  
chemical potentials: 
$\mu_q = \partial \epsilon (\rho_q,\rho_{q'})/ \partial \rho_q$, $q=n,p$, 
 $\epsilon$
 being the energy density. We recall that this quantity contains all 
contributions to the energy per particle (kinetic, potential and 
simmetry energy). 
In Fig.\ref{chem} we report the density dependence of the 
$n,p$ chemical potentials below normal density, where we expect that fragment
formation takes place \cite{baranNPA730}, for
 a system with asymmetry $I = (N-Z)/A = 0.2$ and for two choices 
of the iso-$EoS$.
We refer to an $asy-stiff$ $EoS$ when we consider a potential 
symmetry term
 that is linearly increasing with nuclear density. We refer to 
an $asy-soft$ 
$EoS$ when the potential symmetry term increases up to a 
 saturation around normal density, and then eventually decreases
\cite{ditoro,baran,baranPR}.

Since particles move towards the minimum of the chemical potential, 
it is possible to observe that there exists a density window, roughly
$\rho_0/2 \leq \rho \leq \rho_0$ from Fig.\ref{chem}, where 
neutrons and protons can move in opposite directions: 
protons move 
towards a higher density region while neutrons move towards a lower density
region. 
This phenomenon, called \textit{isospin migration} ~\cite{baran}, 
causes a neck neutron enrichment, due to the density gradient
between the low density neck region and the spectator matter.
We stress that this mechanism is different from the recently
investigated isospin diffusion and N/Z equilibration processes in peripheral
collisions \cite{tsangprl92,souliotisplb588}, that
instead are due essentially to the presence of isospin gradients.

\begin{figure}[t]
\begin{center}
\includegraphics[scale=0.45]{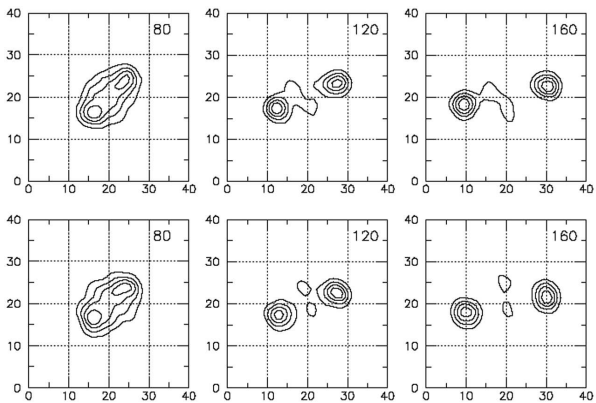}
\caption{Density contour plots on the reaction plane for 
two different events of the reaction $^{58}$Fe + $^{58}$Fe at $47~AMeV$,
 $b_r=0.5$. The box size is $40fm$. The numbers inside the box represent
the time-steps in $fm/c$.}
\label{fedens}
\end{center}
\end{figure}

Reactions have been simulated by considering a 
stochastic extension of the microscopic transport 
equation $BNV$ (Boltzmann-Nordheim-Vlasov), following a 
test-particle evolution 
on a lattice \cite{bertsch,guarnera,greco}.
We consider a
 beam energy of $47~AMeV$, 
and reduced 
impact parameter $b_r \equiv b/b{max} = 0.5$ (semiperipheral). Using the  
$asy-stiff$ $EoS$ 40\% of 
the events produce at least one $IMF$ in the neck region ($ternary~events$). 
In Fig. \ref{fedens}, density contour plots on the reaction plane 
are presented
for two different events coming from the reaction $^{58}$Fe + $^{58}$Fe 
($b_r=0.5,~47~AMeV$).

The first row shows an event in which a fragment forms with a time delay and 
in a space region correlated to a target-like nucleus, unlike 
the event in 
the second row, which shows a rapid fragment formation in a region that 
is not correlated to any spectator remnants. 
Thus fragments are formed according to a variety of mechanisms and we will see
how this influences their properties. 

\begin{figure}[t]
\begin{center}
\includegraphics[scale=0.37]{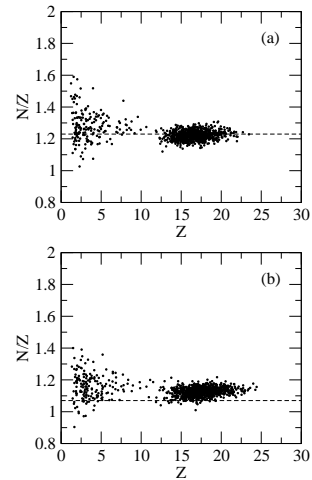}
\caption{Asymmetry vs. charge of each fragment arising from 
the simulation of the reaction $^{58}$Fe + $^{58}$Fe (a) and 
$^{58}$Ni + $^{58}$Ni (b) with an  $asy-stiff~EoS$.
Horizontal dashed lines are the initial asymmetries of the colliding 
systems.}
\label{fenistiff}
\end{center}
\end{figure}

Fig. \ref{fenistiff} reports the $N/Z$ ratio of each fragment vs. the 
charge $Z$ at the freeze-out time, obtained at the intermediate 
impact parameter $b_r=0.5$.

The residual $PLF/TLF$ nuclei (large $Z$ range) show a different behavior
in the two reactions: we note, 
in fact, that 
the points of Fe system are along the dashed line, that 
represents the initial system asymmetry, while for Ni reaction points 
lie above that line. The $IMF$ (low $Z$ range) behavior is similar for the two 
reactions: the points, for both reactions, lie above the 
dashed line, although for the reaction $^{58}$Ni + $^{58}$Ni the
difference between the $N/Z$ of $IMF$'s and large $PLT/TLF$ residues 
seems to be less pronounced.  

For a better understanding of Fig.\ref{fenistiff}, we have to 
consider what happens 
during the pre-equilibrium phase \cite{SobotkaPRC50,Farine,Daniel,Bao}. 
The asymmetry of the di-nuclear neutron-rich
 system changes from 1.23 (initial value) to 1.22 (at t = 100 fm/c, 
instant in which fragments start to form) since 14 neutrons and 11 protons are 
evaporated, while the di-nuclear neutron-poor system changes from 1.07 
to 1.12 as a consequence of a larger proton evaporation, due to the
Coulomb repulsion (it loses 13 protons 
against 12 neutrons), becoming an asymmetric system. We can conclude that:

i) In the neutron rich reaction, the neutron emission due to  
pre-equilibrium goes in the same direction of the neck neutron enrichment, 
caused by the isospin-migration. So finally we observe slightly more symmetric
$PLF/TLF$ residues accompanied by neutron-richer $IMF$'s; 

ii) In the neutron-poor collision, the larger pre-equilibrium proton emission 
enhances the $N/Z$ of the dinuclear composite system. The acquired asymmetry 
is then transferred to the neck region. 

To better disantangle between the two mechanisms, one can study the charge
composition of residues 
distinguishing between binary and ternary events (Table \ref{binter}).

\begin{table} 
\caption{ Asymmetry evolution of the residual nuclei arising 
from binary and ternary events.}
\begin{ruledtabular}
{\footnotesize
\begin{tabular}{c c c c}
\hline
 {} &{} &{} &{}\\[-1.5ex]
$systems$ & $t = 0$ & $t = 100 fm/c$ & $t = 200 fm/c$\\[1ex]
\hline
{} &{} &{} &{}\\[-1.5ex]
$^{58}Fe + ^{58}Fe$ & $1.23$~ & $1.22$~ & $1.23~binary$\\[1ex]
{} & {} & {} & $1.19~ternary$\\[1ex]
\hline
{} &{} &{} &{}\\[-1.5ex]
$^{58}Ni + ^{58}Ni$&$1.07$~&$ 1.12$~&$1.17~binary$\\[1ex]
{} & {} & {} & $1.125~ternary$\\[1ex]
\hline
\end{tabular}}
\label{binter}
\end{ruledtabular}
\end{table}

For both reactions we note that the $N/Z$ ratio of residual $PLF/TLF$
 nuclei in ternary events is lower than the value for binary events, 
 since in the 
latter case the isospin-migration effect does not apply.
The isospin dynamics effect  is rather
evident from the comparison with the asymmetry values at the 
end of the pre-equilibrium
 phase ($t = 100 fm/c$ in the Table). For the $Fe + Fe$ system 
the $N/Z$ of residues changes 
from 1.22 to 1.19, in ternary events; 
for the $Ni + Ni$ reaction this difference is not so evident 
(from 1.12 to 1.125) because the isospin-migration competes with 
proton evaporation.
On the other hand, in binary events, we note the neutron enrichment of 
residues in the $Ni$ reaction, due to a favorite proton pre-equilibrium 
emission. 

\begin{figure}[t]
\begin{center}
\includegraphics [bb = 0 0 1089 496, scale=0.23]{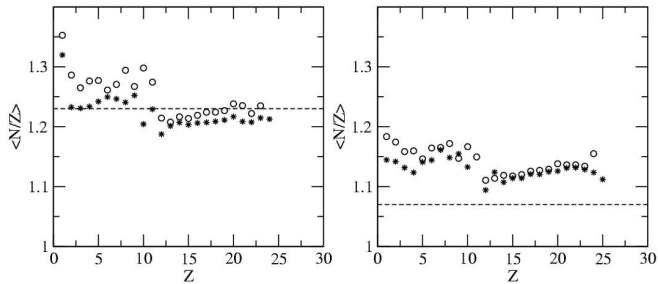}
\caption{Average asymmetry vs. $Z$ for nuclei from the 
reactions $^{58}$Fe + $^{58}$Fe (left panel) and  $^{58}$Ni + $^{58}$Ni
 (right panel) for an $asy-stiff$ (circles) and $asy-soft$ (triangles) $EoS$.
 Horizontal dashed lines represent initial asymmetry of colliding 
systems ~\protect\cite{baran}.}
\label{meanasy}
\end{center}
\vspace*{-10pt}
\end{figure}

In Fig. \ref{meanasy} the average asymmetry of products  
arising from the two reactions is shown for the
two choices of the symmetry energy parameterization,  
$asy-stiff$ and $asy-soft$ $EoS$.

\begin{figure}[t]
\begin{center}
\includegraphics[bb =  16 277 563 552, scale=0.45]{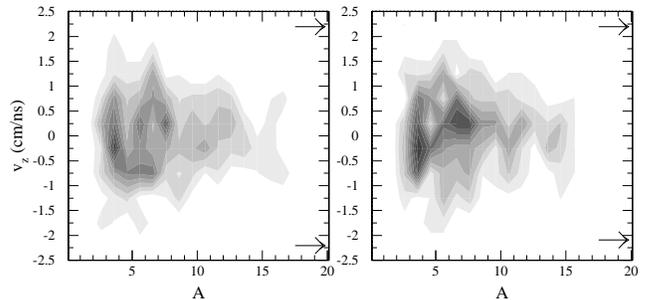}
\caption{Contour plots of the parallel velocity-mass distribution 
for the two systems:
$^{58}$Fe + $^{58}$Fe (left) and $^{58}$Ni +  $^{58}$Ni (right). Arrows
indicate the positions of average PLF and TLF parallel velocity.}
\label{vel_mass}
\end{center}
\end{figure}

\begin{figure}[t]
\begin{center}
\includegraphics[bb = 16 277 563 552, scale=0.45]{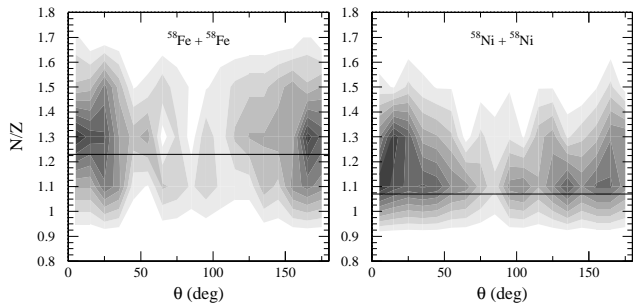}
\caption{Contour plots of the N/Z distribution versus the emission angle.
$Asy-stiff$ choice.}
\label{NZ_angle}
\end{center}
\end{figure}

We note that fragments are more symmetric in 
the asy-soft case. This difference can be explained with the different 
behavior of the chemical potential in the two $EoS$, see Fig. \ref{chem},
that will induce a different isospin-migration effect. 
For an asy-soft $EoS$, the proton chemical potential varies not so much in the 
region where the fragments form (from 0.08 to 0.15 fm$^{-3}$), while for 
neutrons there is  a significant slope (however smaller than 
in the asy-stiff case). 
Therefore neutron enrichment of the neck will still cause an increase of the 
fragment asymmetry, but this increase will be smaller than in the 
asy-stiff case, where protons can even migrate out of the neck region. 
The fact that the $N/Z$ of large fragments is smaller in the asy-soft case
is due to pre-equilibrium effects, since with the asy-soft $EoS$ more neutrons
are emitted due to the more repulsive mean field below normal density, 
\cite{ditoro,baran,baranPR,Bao}.  
On the other hand, the larger  
difference between $IMF$ and $spectator$
fragment asymmetries with the asy-stiff parametrization is a consequence
of a more effective isospin migration, see the recent \cite{baran05}.

It is interesting to look also for some correlations between asymmetry, mass,
velocity and direction of the outgoing fragments. As we will show, in this way
we can even disentangle between features just related to the reaction
mechanism and observables more sensitive to the isospin dynamics and
symmetry energy.

In Fig.\ref{vel_mass} we present the behavior of the velocity along the 
beam direction versus the $IMF$ mass. 
 Lighter fragments are emitted at all
angles and they have larger velocities with respect to more massive $IMF$'s
that are more correlated to the spectator matter and are emitted on longer
time scales. 

The analysis of the $N/Z$ distribution versus the emission angle
(see Fig.\ref{NZ_angle}) reveals
that larger fluctuations are present close to forward and backward angles. 
This indicates that $IMF$'s more correlated to the spectator
matter may become more neutron-rich, since they interact for a longer time
with the system and the isospin transport mechanism becomes more effective.

\begin{figure}[t]
\begin{center}
\includegraphics[bb = 31 275 531 544, scale=0.45]{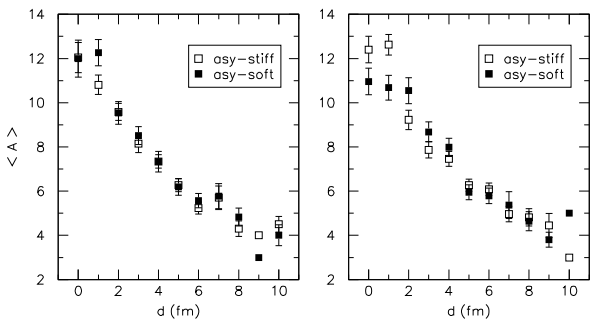}
\caption{Average $IMF$ mass as
a function of the distance from the $PLF-TLF$ axis at the freeze-out time
 for $47~AMeV,~b_r=0.5$ collisions.
Left panel: $Fe+Fe$. Right panel: $Ni+Ni$. Empty squares: $asy-soft$ symmetry
term. Full squares: $asy-stiff$. }
\label{mass_dist}
\end{center}

\end{figure}
\begin{figure}[t]
\begin{center}
\includegraphics[bb = 31 275 531 544, scale=0.45]{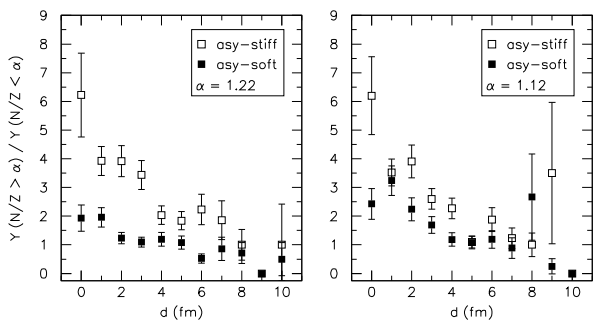}
\caption{Ratio of the $IMF$ yields, with $N/Z$ larger and smaller than the 
value $\alpha$ obtained just after pre-equilibrium emission, as
a function of the distance from the $PLF-TLF$ axis at the freeze-out time
 for $47~AMeV,~b_r=0.5$ collisions.
Left panel: $Fe+Fe$. Right panel: $Ni+Ni$. Empty squares: $asy-stiff$ symmetry
term. Full squares: $asy-soft$. }
\label{asy_dist}
\end{center}
\end{figure}

The presence of correlations between
mass, neutron content and 
kinematical observables can be better evidenced by studying IMF properties
 vs. the ``alignement'', i.e. as a function of the distance $d$ from the
$PLF-TLF$ axis at the freeze-out time.
    
In Fig.\ref{mass_dist} we plot the behavior of the $IMF$ average mass 
 vs. the distance $d$ 
for the n-rich ($FeFe$, left panel)
and the n-poor ($NiNi$, right panel) collision. The calculation is performed
with the two choices of the density dependence of the symmetry term.
In both cases we see a clear increase of the $IMF$ masses with the
fragment ``alignement'', independent of the stiffness of the used 
$Iso-EoS$. This appears a rather general feature of the $neck-fragmentation$
mechanism: light fragments are emitted at earlier times and are not much 
driven by the spectator residues ($PLF/TLF$) \cite{baranNPA730}.
This is also in line with the fact the light fragments may reach
larger velocities (see Fig.\ref{vel_mass}).

As a consequence we would expect a different amount of isospin migration
 vs. the $IMF$ alignement, and now such correlation should be $Iso-Eos$
dependent. In Fig.\ref{asy_dist} we report the ratio of the $IMF$ yields
with the $N/Z$ larger and smaller than the value, $\alpha$, reached just after
pre-equilibrium emission (see Table I), plotted vs. the
alignement distance $d$ at freeze out. For both systems we have a nice
increase of the neutron enrichment with the alignement. As expected from the 
previous discussion on the physics of the $isospin~migration$, the effect
is more evident in the $asy-stiff$ choice.

We stress again that $neck-IMF$'s always present a neutron enrichment,
{\it even in the case of a n-poor system}. The latter paradox is due
to the pre-equilibrium isospin dynamics. 
In fact, as anticipated above, 
due to pre-equilibrium emission, the system will loose some protons 
and acquire a $N/Z$ larger than the initial one.  Then, before the di-nuclear
system reseparates, the neutron excess is transferred to the neck region
that is at lower density. 
Both effects, fast proton emission and neutron transfer, are connected to
the symmetry term of the $EoS$.
Some evidence has been found in 
recent data on $^{58}$Ni induced fragmentation 
\cite{milazzo,GingrasPRC65,MoustbchirNPA739}.

In conclusion, 
 we have shown that isospin dynamic processes appear 
even in systems with initially uniform spatial asymmetry distribution, 
such as $^{58}$Fe + $^{58}$Fe and  $^{58}$Ni + $^{58}$Ni.
From a chemical point of view,
we do not expect $N/Z$ gradients which can induce asymmetry 
variations. These variations are instead caused by density gradients
during the reaction dynamics since
the symmetry 
term of the $EoS$  introduces a different behavior of the chemical 
potentials for neutrons and protons with respect to density. So, when the 
collision happens, the spatial distribution of the isoscalar density will 
induce 
variations even in the isovector density.

We have revealed an interesting correlation, typical of the 
$neck-fragmentation$ mechanism, between $IMF$ masses and corresponding 
alignement to the axis joining the $PLF-TLF$ residues. This suggests a 
time-hierarchy in the mid-rapidity fragment production with the lighter
clusters formed at earlier times. When we combine to the isospin migration 
dynamics, an $Iso-Eos$ sensitive observable results to be the correlation 
between the neutron excess of $IMF$'s and the relative alignement.

The reactions $^{58}$Fe + $^{58}$Fe and $^{58}$Ni + $^{58}$Ni 
have been studied experimentally, focussing on central collisions 
and adopting a statistical description of multifragmentation 
in \cite{shetty,shetty2}, or in semi-peripheral collisions in \cite{milazzo}.
Asymmetry 
effects, as the ones described here,  
have been noted in \cite{milazzo}, even in the second system.  
In this paper we suggest new interesting 
 correlation analyses that can be pursued selecting semi-peripheral
events and that appear particularly appropriate in 
order to investigate the isovector structure of the $EoS$ as well as to 
shed some lights 
on the mechanism responsible for fragment production.

\end{document}